\documentclass[conference]{IEEEtran}
\IEEEoverridecommandlockouts
\usepackage{cite}
\usepackage{amsmath,amssymb,amsfonts}
\usepackage{algorithmic}
\usepackage{graphicx}
\usepackage{algorithm}
\usepackage{textcomp}
\usepackage{xcolor}
\usepackage{amsmath,amsfonts,amsthm,bm} %
\usepackage{cuted}
\usepackage{multirow}
\usepackage{float} 
\newcommand{\tabincell}[2]{\begin{tabular}{@{}#1@{}}#2\end{tabular}}

\def\BibTeX{{\rm B\kern-.05em{\sc i\kern-.025em b}\kern-.08em
    T\kern-.1667em\lower.7ex\hbox{E}\kern-.125emX}}
    
\begin{document}

\title{Pilot Aided Channel Estimation for AFDM in Doubly Dispersive Channels\\
	\thanks{This work is funded by Guangdong Natural Science Foundation
		under Grant 2019A1515011622.}
	
}

\author{\IEEEauthorblockN{Haoran Yin, Yanqun Tang*}
	\IEEEauthorblockA{
		\textit{School of Electronics and Communication Engineering, Sun Yat-sen University, China}
		\\
		email: yinhr6@mail2.sysu.edu.cn, tangyq8@mail.sysu.edu.cn}
}

\maketitle

\begin{abstract}
\textbf{Affine frequency division multiplexing (AFDM) is a multi-chirp waveform and a promising solution for ultra-reliable communication under doubly dispersive channels. In this paper, we propose two pilot aided channel estimation schemes for AFDM, named single pilot aided (SPA) and multiple pilots aided (MPA) respectively. Pilot, guard and data symbols in the discrete affine Fourier transform (DAFT) domain are arranged appropriately at the transmitter.  While at the receiver, channel estimation is performed with the aid of an estimation threshold and a mapping table. The bit error performance of AFDM applying the proposed channel estimation schemes shows only marginal loss compared to AFDM with ideally known channel state information. Moreover, extensions of the SPA scheme to multiple-input multiple-output (MIMO) and multi-user uplink/downlink are presented. }
\end{abstract}


\section{Introduction}
The beyond 5G / 6G wireless networks are expected to meet the requirements of reliable, low latency and wide range communication under high-speed scenarios. It is a huge challenge since the channel is no longer linear time-invariant as 4G systems. Under this time dispersive channels, the orthogonality of orthogonal frequency division multiplexing (OFDM) is damaged greatly by the doppler shift, resulting in intolerable performance loss. Orthogonal chirp division multiplexing (OCDM)\cite{b3}, which is based on the discrete Fresnel transform, outperforms OFDM in doubly dispersive channels. However, OCDM cannot achieve full diversity due to its unchangeable parameters. Orthogonal time frequency space (OTFS) modulation\cite{b7}, a recently proposed two-dimensional modulation scheme, has shown the potential of tackling the dynamics of doubly dispersive channels with symbols multiplexed in the delay-doppler (DD) domain via symplectic finite Fourier transform. Given the sparse and stable property of the representation of channel characteristics, OTFS is resilient to delay-doppler shifts and outperforms OFDM significantly.

Affine frequency division multiplexing (AFDM), a recently discovered waveform, always attains full diversity due to parameters adjustment according to the DD profile of the channel\cite{b1}. Symbols in AFDM are multiplexed on a set of orthogonal chirps which are tuned to adapt to the doubly dispersive channel characteristics, enabling a full  delay-doppler representation of the channel in discrete affine Fourier transform (DAFT) domain \cite{b5}. Results in\cite{b1} \cite{b2} show that AFDM has similar outstanding performance just as OTFS but with lower complexity and advantage on less channel estimation overhead.

A fundamental building block in wireless communication systems is channel estimation, which acquires the accurate channel state informantion (CSI) during the transmission. The performance of detection at the receiver is deeply determined by how accurate the CSI is estimated. Embedded pilot channel estimation is mentioned in \cite{b2} with less explanation and no numerical results. To the best of our knowledge, this is the first work introducing the pilot aided channel estimation with systematic analysis in AFDM systems, regardless of single antenna or multiple antennas.

In this paper, we propose two pilot aided channel estimation schemes for AFDM, named single pilot aided (SPA) and multiple pilots aided (MPA) respectively. Channel estimation is performed with the aid of an estimation threshold and a mapping table that can extract the CSI perfectly from the received pilot symbols. Extensions of the SPA channel estimation scheme to multiple-input multiple-output (MIMO) and multi-user uplink/downlink scenarios are discussed. The simulation results show that,  the bit error performances based on the proposed  schemes of SPA and MPA achieve that of  AFDM with ideal CSI. The rest of this paper is organized as follows. Section II reviews basic concepts of AFDM, which lays the foundations for the development of channel estimation schemes in Section III. Numerical results are presented in Section IV, followed by the conclusions in Section V.

\section{ Basic Concepts of AFDM}
\begin{figure}[htbp]
	\centering
	\includegraphics[width=0.48\textwidth,height=0.10\textwidth]{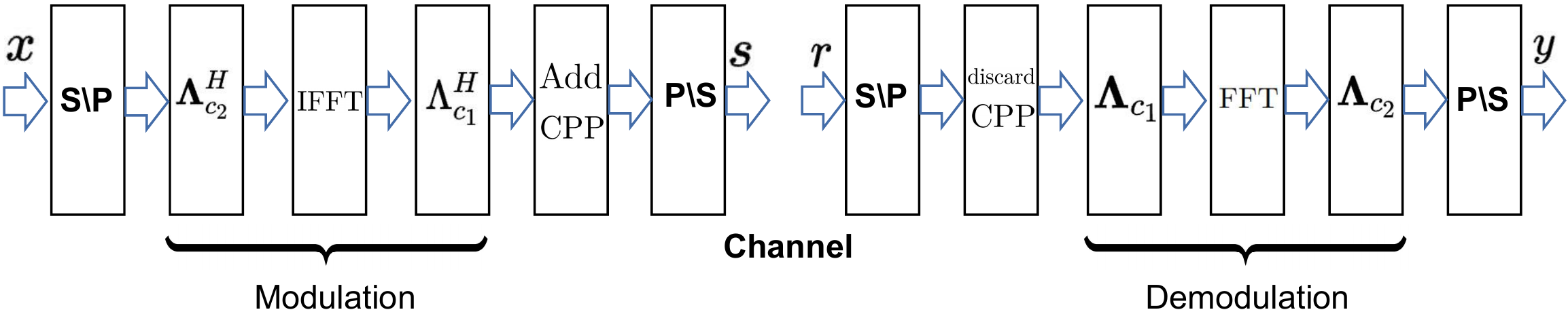}
	\caption{ AFDM modulation/demodulation block diagrams}
	\label{2-1}
\end{figure}

In this section, we review the basic concepts of AFDM from\cite{b1}\cite{b2},  Fig. \ref{2-1} shows the AFDM modulation/demodulation block diagrams.

Let $\mathbf{x}$ denote a vector of $N$ quadrature amplitude modulation (QAM) symbols that reside on the DAFT domain. The $N$ points inverse DAFT (IDAFT) is performed to map $\mathbf{x}$ to the time domain as

\begin{equation}
s_{n}=\frac{1}{\sqrt{N}} \sum_{m=0}^{N-1} x_{m} e^{i 2 \pi\left(c_{2} m^{2}+\frac{1}{N} m n+c_{1} n^{2}\right)}
\label{eq1}
\end{equation}
with $n=0, \cdots, N-1$, $N$ denoting the number of subcarrier, and $\boldsymbol{\Lambda}_{c}=\operatorname{diag}\left(e^{-i 2 \pi c n^{2}}, n=0,1, \ldots, N-1\right)$. Before transmitting $\mathbf{s}$ into the channel, a $chirp-periodic$ prefix (CPP) should be added
with a length which is any integer greater than or equal to the value in samples of the maximum delay spread of the wireless channel.

After transmitting through the channel, serial to parallel, discarding  CPP and $N$ points DAFT are performed. Thus, the received time domain samples $\mathbf{r}$ are transformed to DAFT domain samples $\mathbf{y}$ with
\begin{equation}
y_{m}=\frac{1}{\sqrt{N}} \sum_{n=0}^{N-1} r_{n} e^{-i 2 \pi\left(c_{2} m^{2}+\frac{1}{N} m n+c_{2} n^{2}\right)}
\end{equation}
where $m=0,...,N-1$. As proven in \cite{b1}, the input-output relation between $\mathbf{y}$ and $\mathbf{x}$ satifies
\begin{equation}
y_{p}=\sum_{i=1}^{P} h_{i} e^{i \frac{2 \pi}{N}\left(N c_{1} l_{i}^{2}-q l_{i}+N c_{2}\left(q^{2}-p^{2}\right)\right)} x_{q}+\tilde{w}, 0 \leq p \leq N-1
\label{eq3}
\end{equation}
where the noise in DATF domain $\tilde{w} \sim \mathcal{C} \mathcal{N}\left(0, N_{0}\right)$, $P \geq 1$ is the number of paths, $q=(p+\operatorname{loc}_{i})_{N}$,   $\operatorname{loc}_{i}\triangleq\alpha_{i}+2 N c_{1} l_{i}$, $\alpha_{i} \in\left[-\alpha_{\max }, \alpha_{\max }\right]$ is the integer part of doppler shift  corresponding to $i$-th path normalized with the subcarrier spacing, while in this paper we assume the fractional part is zero and $\alpha_{\max }$ is the maximum doppler shift. $l_{i}$ denotes the delay spread normalized with sample period, $c_{1}=\frac{2 \alpha_{\max }+1}{2 N}$ and $c_{2}$ is either an
arbitrary irrational number or a rational number sufficiently smaller than $\frac{1}{2N}$.
Equation (\ref{eq3}) can be denoted in matrix form as 
\begin{equation}
	\mathbf{y}=\mathbf{H}_{\mathrm{eff}} \mathbf{x}+\tilde{\mathbf{w}}
	\label{eq4}
\end{equation}
where the effective channel matrix in DAFT domain $\mathbf{H}_{\mathrm{eff}}=\boldsymbol{\Lambda}_{c_{2}} \mathbf{F} \boldsymbol{\Lambda}_{c_{1}} \mathbf{H} \boldsymbol{\Lambda}_{c_{1}}^{H} \mathbf{F}^{H} \boldsymbol{\Lambda}_{c_{2}}^{H}$, with $\mathbf{F}$ denoting the  discrete Fourier transform (DFT) matrix, $\mathbf{H}$ being the matrix representation of the channel, noise vector $\tilde{\mathbf{w}} \sim \mathcal{C} \mathcal{N}\left(\mathbf{0}, N_{0} \mathbf{I}\right)$.

 \begin{figure}[b]
	\centering
	\includegraphics[width=0.48\textwidth,height=0.45\textwidth]{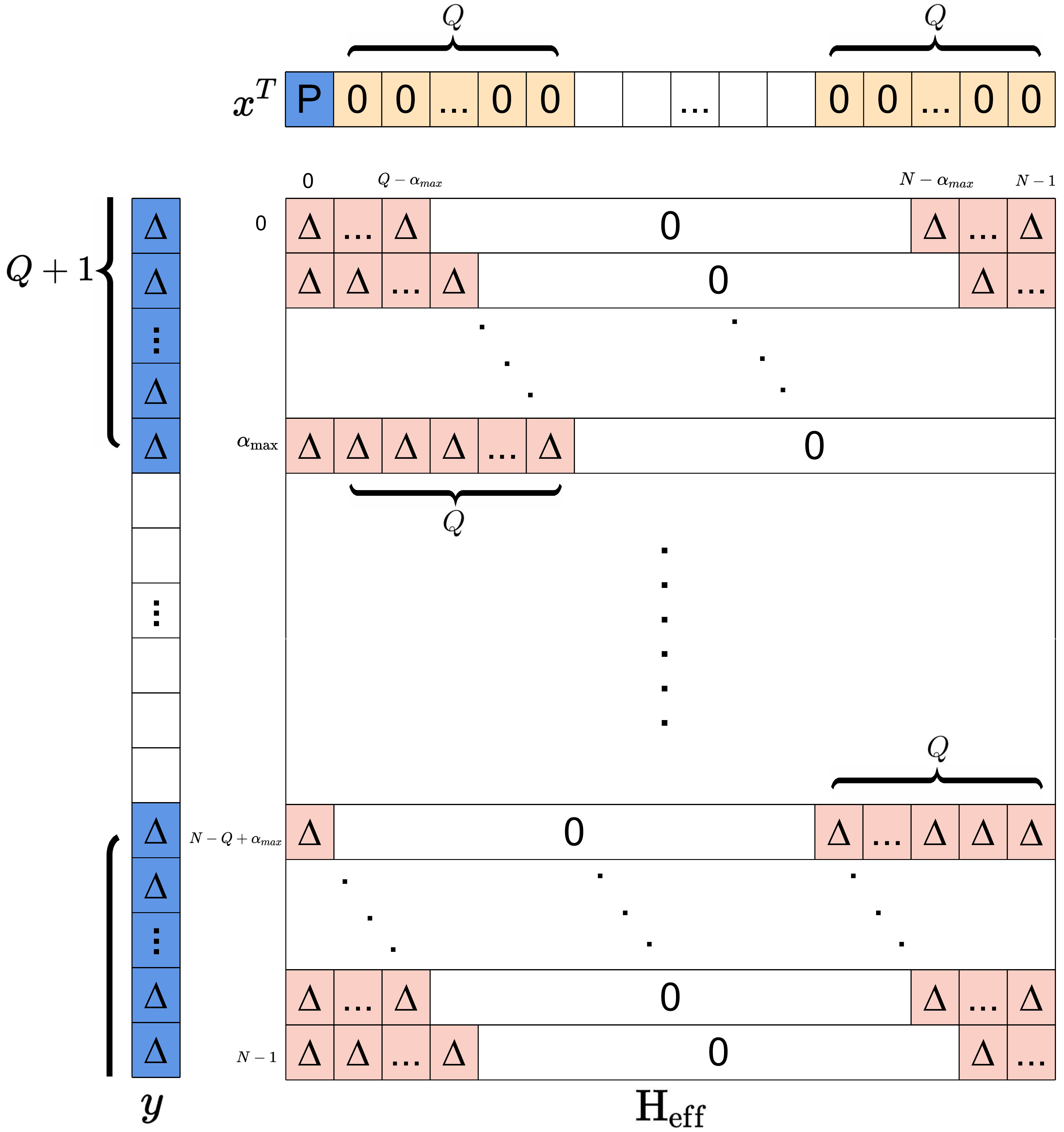}
	\caption{ Effective channel matrix and SPA scheme. (‘p’:pilot, ‘0’:guard, ‘$\Delta$’:non-zero value)}
	\label{3-1}
\end{figure}

\section{Pilot Aided Channel Estimation }
In this section, we propose SPA and MPA channel estimation schemes by exploring the characteristics of the effective channel matrix $\mathbf{H}_{\mathrm{eff}}$. Both of the schemes are performed in the DAFT domain. Based on the received pilots, an estimation threshold and a mapping table are used to extract channel state information. Finally, the SPA scheme is extended to MIMO and multi-user uplink/downlink scenarios. 

\subsection{SPA and MPA schemes}
In AFDM, there are $Q+1$ possible non-zero entries representing $Q+1$ paths in each row and column in the effective channel matrix $\mathbf{H}_{\mathrm{eff}}$, as shown by Fig.\ref{3-1}, where $Q \triangleq 2 l_{\max } \alpha_{\max }+2 \alpha_{\max }+l_{\max }$, $l_{max}$ is the maximum delay spread normalized with sample period. 
It is important to notice that the indexes of possible non-zero entries are circulant in both row and column directions, which indicates that each row and column contains all the CSI.

Following the  above analysis, a single pilot combined with guard symbols can be applied to perform channel estimation as presented by Fig.\ref{3-1}, with the assumption that the pilot is placed in the $0$-th position. Here we elaborate how to arrange the guard symbols to protect the pilot symbols at the receiver from the interference of the other symbols. Firstly, identify the indexes of received pilot symbols, which equal to sorting out the indexes of possible non-zero entries in the 0-th column from $\mathbf{H}_{\mathrm{eff}}$. In this case, it is [$0, \alpha_{max}$] and [$N-Q+\alpha_{max}, N-1$]. Secondly, find out all the other symbols that will affect the received pilot symbols and set them to be zeros. After that, the DAFT domain representation of the channel can be obtained at the receiver. The left blank slots in $\mathbf{x}$ can be further explored. If data is placed to the blank slots,  the SPA scheme degenerates into the embedded pilot scheme mentioned in \cite{b2}.

\begin{figure}[htbp]
	\centering
	\includegraphics[width=0.49\textwidth,height=0.18\textwidth]{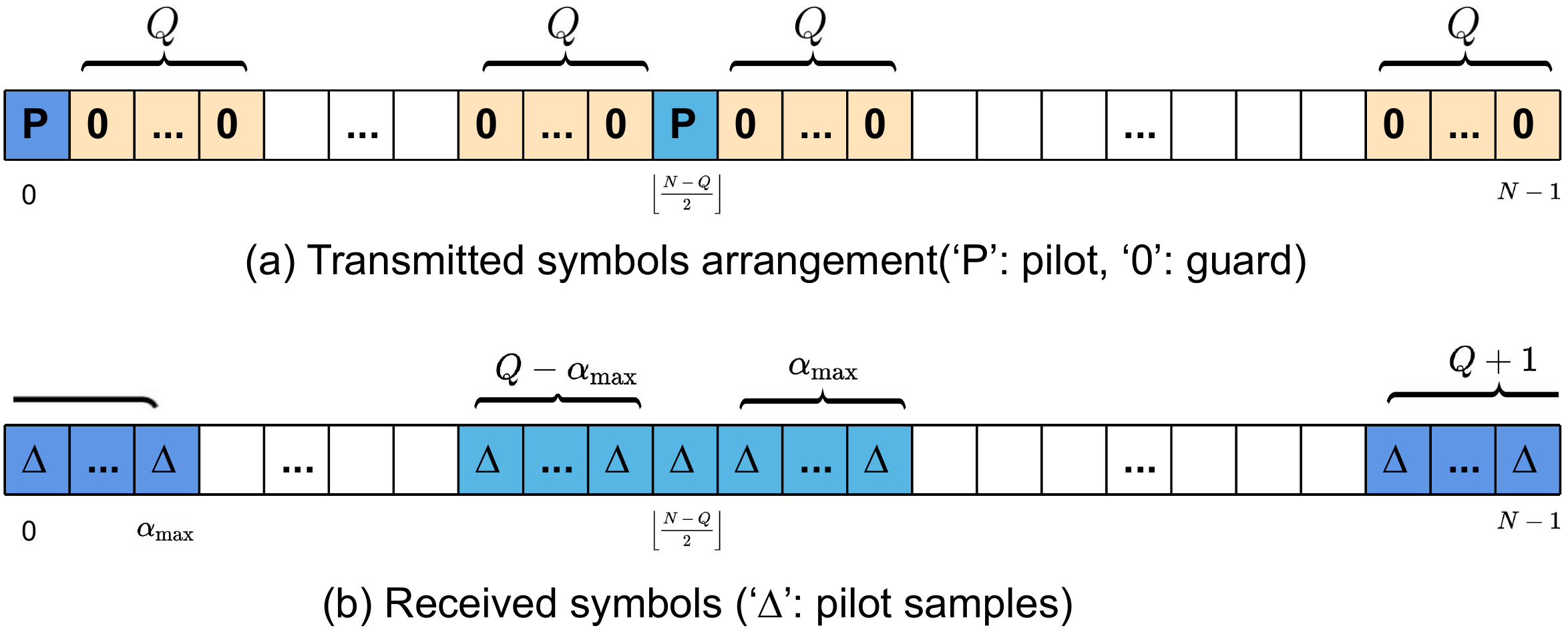}
	\caption{ Two pilots aided channel estimation.}
	\label{3-2}
\end{figure}

Although the guard symbols in SPA can protect the received pilot symbols from other symbols, its estimation accuracy degrades under high noise condition. In order to cope with the interference from noise, multiple pilots can be inserted in the blank slots since all the columns in $\mathbf{H}_{\mathrm{eff}}$ contain the same CSI. With guard symbols arrangement similar to SPA,  different groups of received pilot symbols corresponding to different pilots will not interfere with each other in MPA. Fig.\ref{3-2} displays an example of two pilots aided channel estimation. Pilot diversity gain can be extracted to enhance the channel estimation accuracy with the fact that the noise contained in different received symbols is independent identically distributed\cite{b8}. In order to perform channel estimation successfully, $N$ in SPA and MPA should satisfy

\begin{equation}
 N_{p}(2l_{\max }\alpha_{\max }+2\alpha_{\max }+l_{\max }+1)\le N
\end{equation}
where $N_{p}$ denotes the number of used pilots.

\subsection{Extract CSI from received pilots}
At the receiver, CSI can be estimated by making use of the received pilot symbols. To identify the existence of a path, we introduce a threshold $\zeta$ normalized with noise variance $N_{0}$. If the energy of the received pilot symbol exceeds $\zeta$, then we get a positive detection to the corresponding path, otherwise a negative detection will be obtained. The performance of threshold aided channel estimation scheme is deeply affected by the threshold value as the simulation results shown in section IV. If the threshold is set too high, miss detection of the existing path may occur, if the threshold is set too low, false detection may happen on a path that doesn't exist. Both of the cases will deteriorate the accuracy of channel estimation, along with the detection performance degradation. 

After sorting out the received pilot symbols with energy exceed to the estimation threshold $\zeta$, the CSI in terms of DD profile can be extracted directly with a mapping table, which 
essentially provides the channel conversion from DAFT domain to DD domain.
For example, if a pilot $x_{p}$ is placed in the 0-th slot as Fig.\ref{3-1}, then
 the mapping relationship between the received pilot symbol $y_{m}$, $m\in [0, \alpha_{max}]\cup [N-Q-\alpha _{max}, N-1]$, and the path with integer delay $l_{i}$, doppler shift $\alpha_{i}$, and complex gain $h_{i}$ is shown in Table I, where $E \triangleq e^{i \frac{2 \pi}{N}\left(N c_{1} l_{i}^{2}-N c_{2} m^{2}\right)}$. The mapping relationship for pilots inserted in different slots in the transmitted AFDM frame can be easily obtained with a little adjustment of Table I according to (\ref{eq3}).

 Although thorough awareness of the DD profile of the channel is not necessary for equalization and detection in AFDM system\cite{b10}, it is extremely important and useful in the field of integrated sensing and communication (ISAC). In the ISAC systems, the delay and doppler reflect the distance and velocity information of objects respectively and AFDM is a promising waveform \cite{b9}.

\begin{table}[]
	\renewcommand\arraystretch{1.3}
	\centering
	\caption{From DAFT domain to DD domain}
	\label{t1}
	\begin{tabular}{|c|c|c|c|}
		\hline
		$\bm{m}$& $\bm{l_{i}}$ &$\bm{\alpha_{i}}$&$\bm{h_{i}}$\\ \hline
		 $\left[ 0,\alpha_{max}\right]$ &0 &$-m$&\multirow{3}{*}{\tabincell{c}{\\ \\$\frac{y[m]}{E\cdot x_{p}} $}} \\ \cline{1-3} 
		$\left[ N-\alpha_{max}, N-1\right]$& 0 &$N-m$&\\   \cline{1-3} 
		\tabincell{c}{$[ N-\alpha_{max}-2Nc_{1}d$, \\ $N-\alpha_{max}-2Nc_{1}(d-1)-1 $] \\ $d=1,...,l_{max}$ } & $d$ &$N-m-2Nc_{1}d$&\\ \hline
	\end{tabular}
\end{table}

\subsection{MIMO scenario}
In the sequel, we apply the SPA scheme in the scenario of MIMO. For simplicity, we discuss the case of single pilot embedded in the data with the assumption that the $l_{max}$ and $\alpha_{max}$ are identical between the antennas at the transmitter (Tx) and antennas  at the receiver (Rx).

\begin{figure}[htbp]
	\centering
	\includegraphics[width=0.48\textwidth,height=0.35\textwidth]{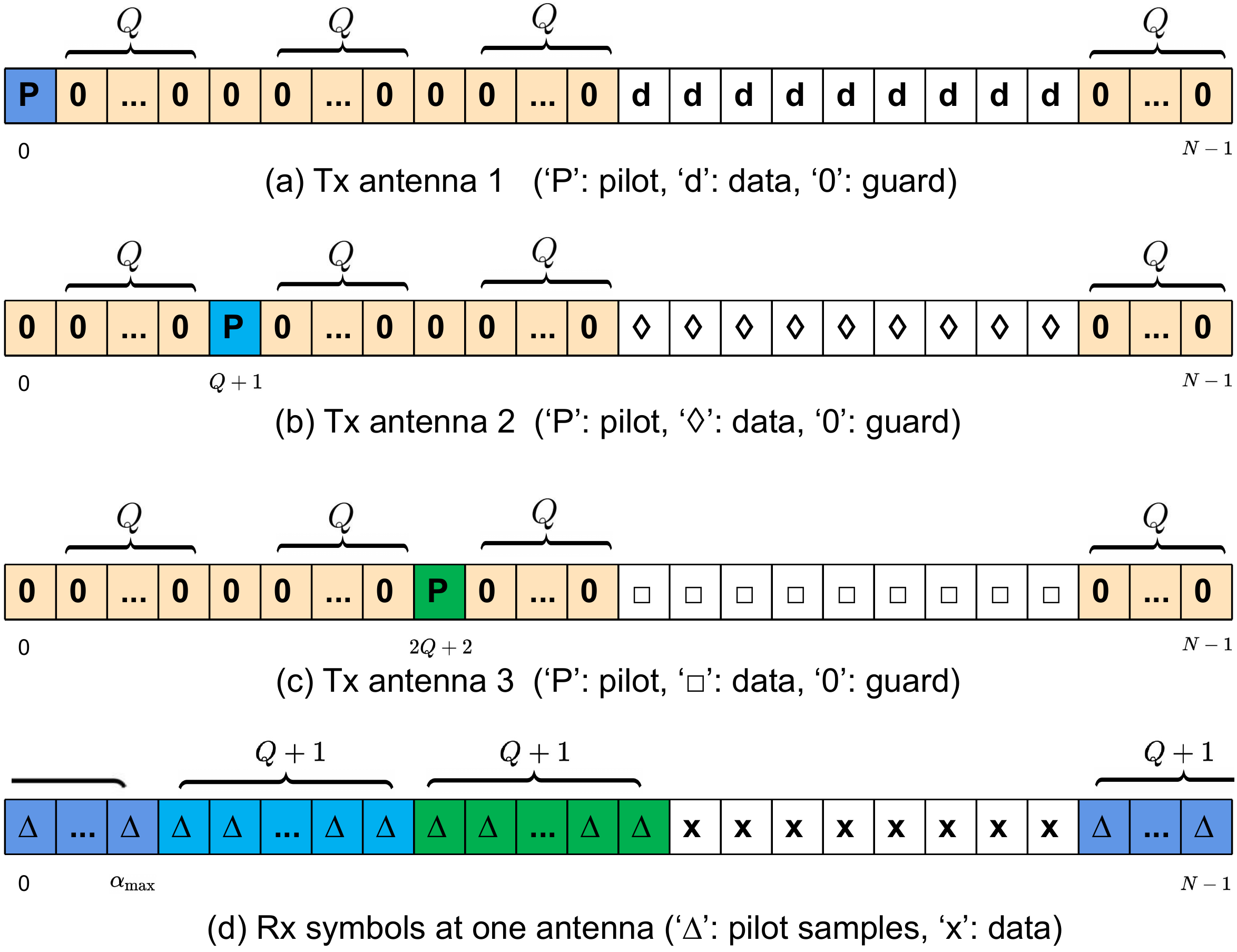}
	\caption{ $3 \times 3$ MIMO-AFDM system.}
	\label{3-4}
\end{figure}

Let $N_{t}$ and $N_{r}$ denote the number of antennas in the transmitter and receiver respectively. In the transmitter, each of Tx antennas places its pilot in a specific location surrounded by guard symbols. While in the receiver, each of the Rx antennas explores the $N_{t}$ groups of received pilot symbols for channel estimation. Inspired by the mentioned channel estimation schemes above, we propose the following symbol arrangement for arbitrary $N_{t}\geq1$ and $N_{r}\geq1$ MIMO-AFDM system:
\begin{equation}
	x^{n_{t}}[m]=\left\{\begin{array}{ll}
		x_{p}^{n_{t}} & m = (Q+1)(n_{t}-1)\\
		x_{d}^{n_{t}} & m = (Q+1)N_{t}, ..., N-Q-1 \\
		0 & otherwise \\
	\end{array}\right.
\end{equation}
where $x_{p}^{n_{t}}$ and $x_{d}^{n_{t}}$ denote the pilot and data symbols of $n_{t}$-th Tx antenna
respectively. An example of $3\times3$ MIMO-AFDM system is shown in Fig.\ref{3-4}.  We can see that the pilot symbols of the Tx antennas are sufficiently separated so that they do not interfere with each other at the Rx antennas.

The pilot and guard overhead of an $N_{t} \times N_{r}$ MIMO-AFDM for channel estimation can be expressed as 
\begin{equation}
\begin{aligned}
	O_{AFDM} &=(N_{t}+1)Q+N_{t}\\
	&=2N_{t}l_{max}\alpha _{max}+2l_{max}\alpha _{max}+2N_{t}\alpha _{max}+\\
	&\quad   N_{t}l_{max}+2\alpha _{max}+l_{max}+N_{t}
\end{aligned}
\end{equation}
while the overhead of $N_{t} \times N_{r}$ MIMO-OTFS system is\cite{b6} 
\begin{equation}
\begin{aligned}
	O_{OTFS} &= \big( (N_{t}+1)l_{max}+N_{t}\big )(4\alpha_{max}+1)\\
	&=4N_{t}l_{max}\alpha _{max}+4l_{max}\alpha _{max}+4N_{t}\alpha _{max}+\\
	&\quad   N_{t}l_{max}+l_{max}+N_{t}
\end{aligned}
\end{equation}
By comparing with the overheads of AFDM and OTFS systems, we can conclude that AFDM maintains its advantage on OTFS in terms of overhead for channel estimation in MIMO system.

\subsection{Multiuser scenario}
As mentioned above, it is essential that we adopt the pilot aided estimation scheme in the scenarios of multiuser uplink and downlink. Thus, we also consider the single pilot embedded in the data, and assume that the $l_{max}$ and $\alpha_{max}$ are identical between the base station (BS) and different users.

\begin{figure}[htbp]
	\centering
	\includegraphics[width=0.49\textwidth,height=0.34\textwidth]{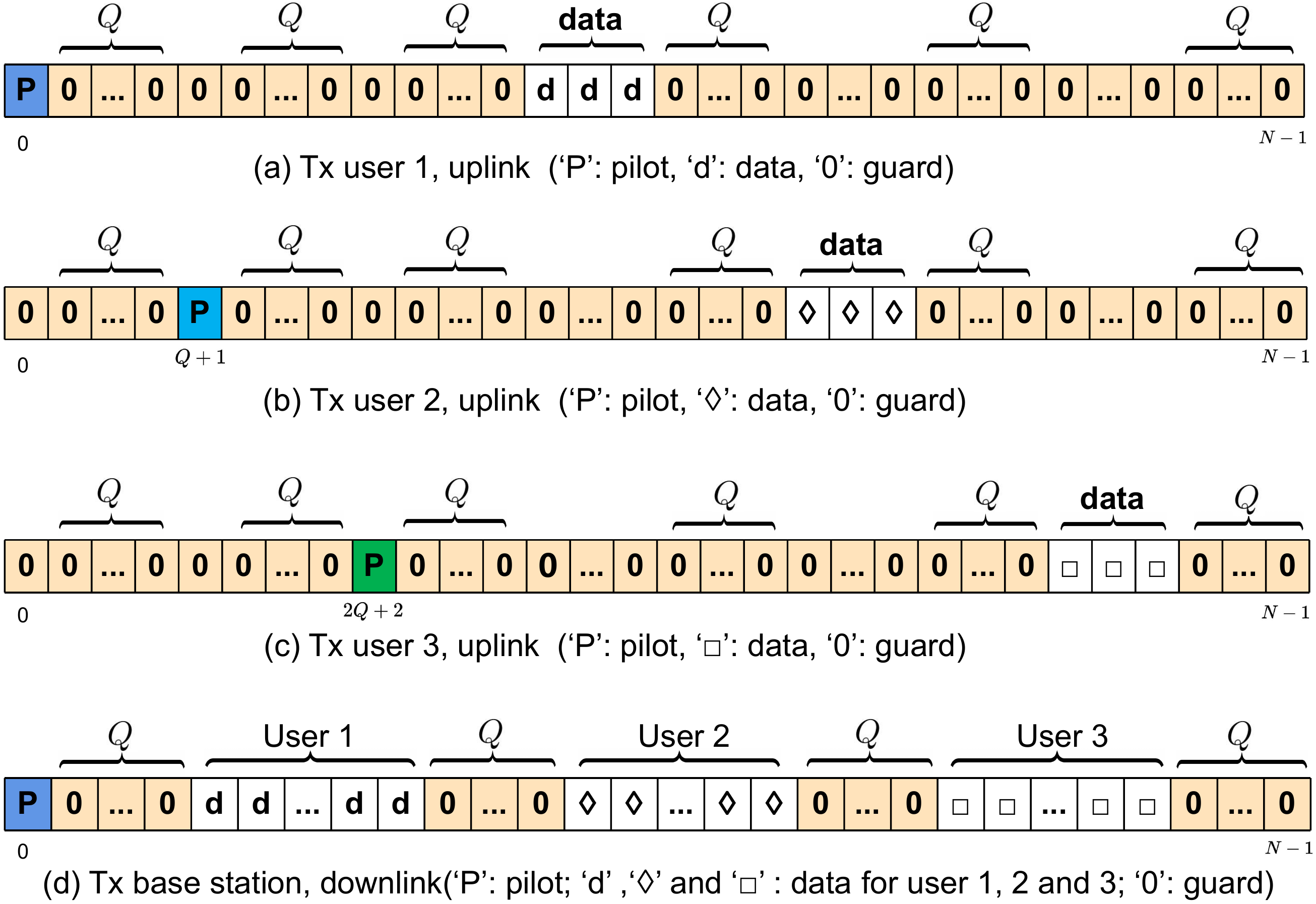}
	\caption{ Single antenna multiuser AFDM system.}
	\label{3-5}
\end{figure}

 $1) Uplink:$ The arrangement of pilots for different users is similar to the MIMO case. Each user occupies only a non-overlapping portion of the rest of the slots for data transmissions.
Zero guard symbols are inserted between the data symbols of different users to avoid inter-user interference (IUI). An example of three users case is shown in Fig.\ref{3-5}(a)(b)(c).

 $2) Downlink:$
During the downlink transmission, only one pilot is needed in the Tx antenna of the BS since it can be used by all the users to estimate the channel from itself to the BS. Guard symbols are also required to avoid IUI, as an example of single-antenna BS with three users shown in Fig.\ref{3-5}(d).

\section{Simulation Results}
We illustrate the performance in terms of bit-error-rate (BER) of the single-input single-output (SISO) and MIMO AFDM system applying the proposed channel estimation schemes. The following system parameters are adopted: carrier frequency of 4 GHz, sub-carrier spacing of 444 Hz, quadrature phase shift keying (QPSK) signaling, linear minimum
mean square error (LMMSE) detector, energy ratio of pilot to noise SNRp$= \left|x_{p}\right|^{2} / N_{0}$ and average energy ratio of data to noise SNRd $= \mathbb{E}\left(\left|x_{d}\right|^{2}\right) / N_{0}$. Extended Vehicular A model \cite{b11} is applied, and each delay tap has a single doppler shift generated by using Jakes's formula, i.e.,$\alpha_{i}=\alpha_{\max } \cos \left(\theta_{i}\right)$, where $\alpha_{max}=4$ corresponding to maximum UE speed of 480 Kmph, and $\theta_{i}$ is uniformly distributed over $[-\pi, \pi]$.

\begin{figure}[hbp]
	\centering
	\includegraphics[width=0.48\textwidth,height=0.38\textwidth]{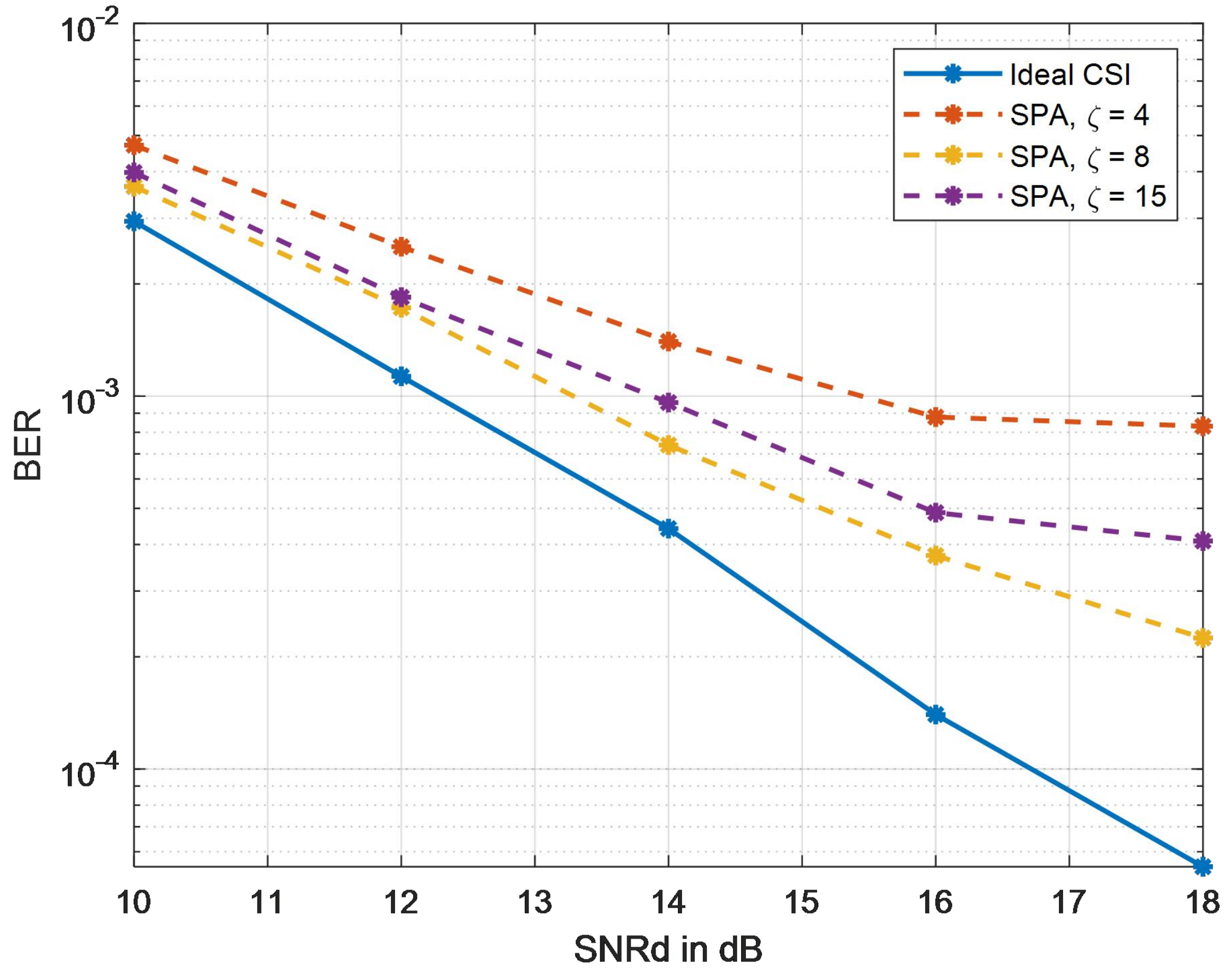}
	\caption{BER versus SNRd for different estimation thresholds, SNRp = 30 dB, $N$ = 256.}
	\label{4-1}
\end{figure}

Fig.\ref{4-1} shows the BER versus SNRd for AFDM with estimation threshold $\zeta_{1} = 4$, $\zeta_{2}=8$ and  $\zeta_{3}=15$, while SNRp $=30$dB. AFDM with ideal CSI is given as a reference. With single pilot scheme adopted, we observe that the estimation threshold $\zeta_{2}=8$ enjoys better performance in comparison with $\zeta_{1}=4$ and $\zeta_{3}=15$. This is because the probability of false detection for inexistent paths is the highest in the case of $\zeta_{1}=4$, and the probability of miss detection for existing paths is the highest in the case of $\zeta_{3}=15$, both of which result in BER performance degradation. The results prove that the choice of estimation threshold can affact greatly the accuracy of channl estimation just as the analysis in Section III, and there exists an optimal estimation threshold under certain condition.

\begin{figure}[htbp]
	\centering
	\includegraphics[width=0.49\textwidth,height=0.39\textwidth]{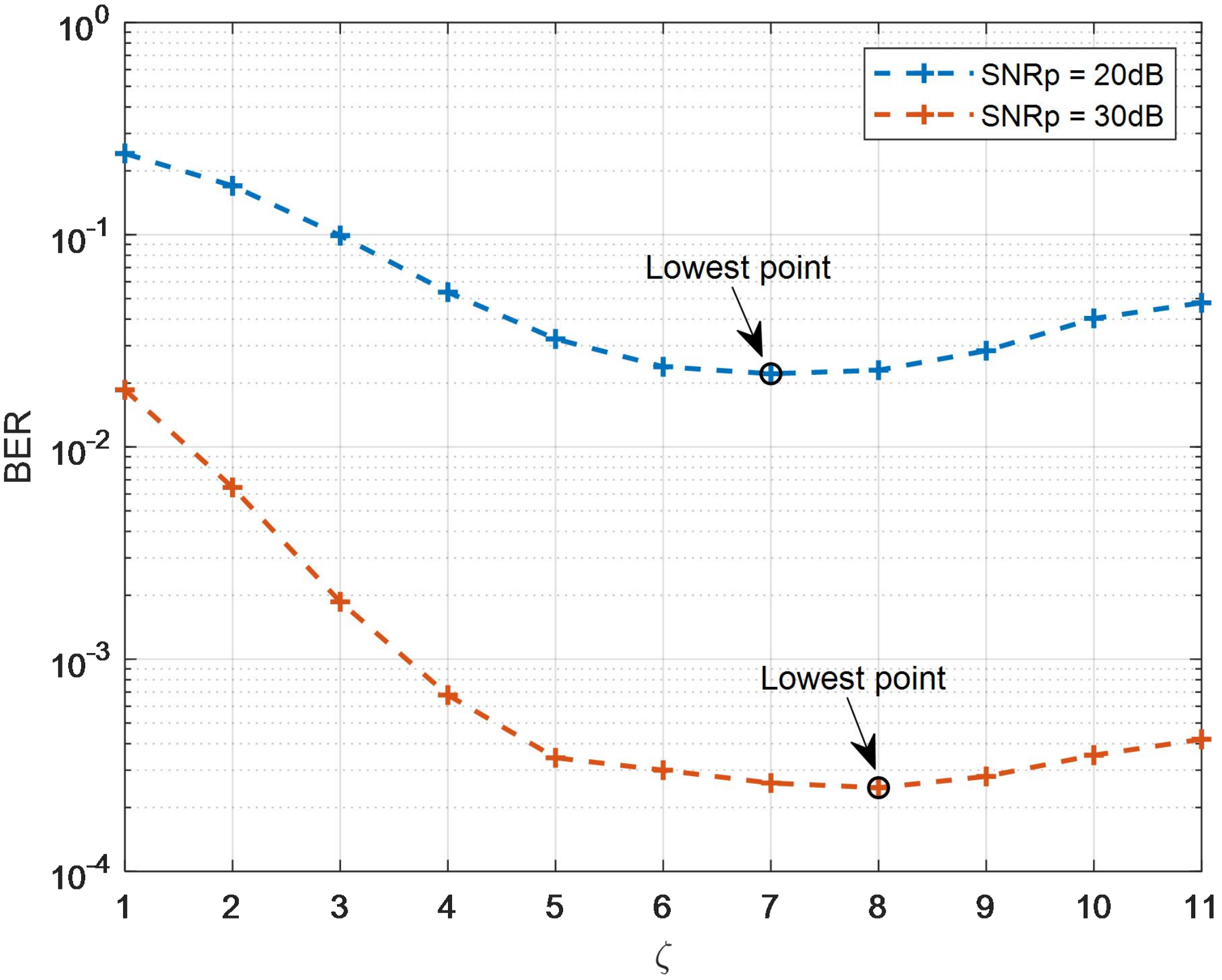}
	\caption{BER versus different estimation thresholds, SNRd = 18 dB, $N$ = 256.}
	\label{4-2}
\end{figure}

We next investigate the effect of SNRp on the optimal channel estimation threshold with SPA  scheme. We fix the SNRd to 18 dB, and Fig.\ref{4-2} shows the BER versus different estimation thresholds with SNRp = 20dB and SNRp = 30dB. We observe that there exist an optimal estimation threshold for each case, and the optimal choice in SNRp = 20dB is 7, while that in SNRp = 30dB is 8. This means, with the energy of the received pilot symbols stronger, a larger estimation threshold should be applied to avoid false detection, while a smaller estimation threshold should be used to avoid miss detection with the energy of the received pilot symbols become weaker.

\begin{figure}[htbp]
	\centering
	\includegraphics[width=0.50\textwidth,height=0.39\textwidth]{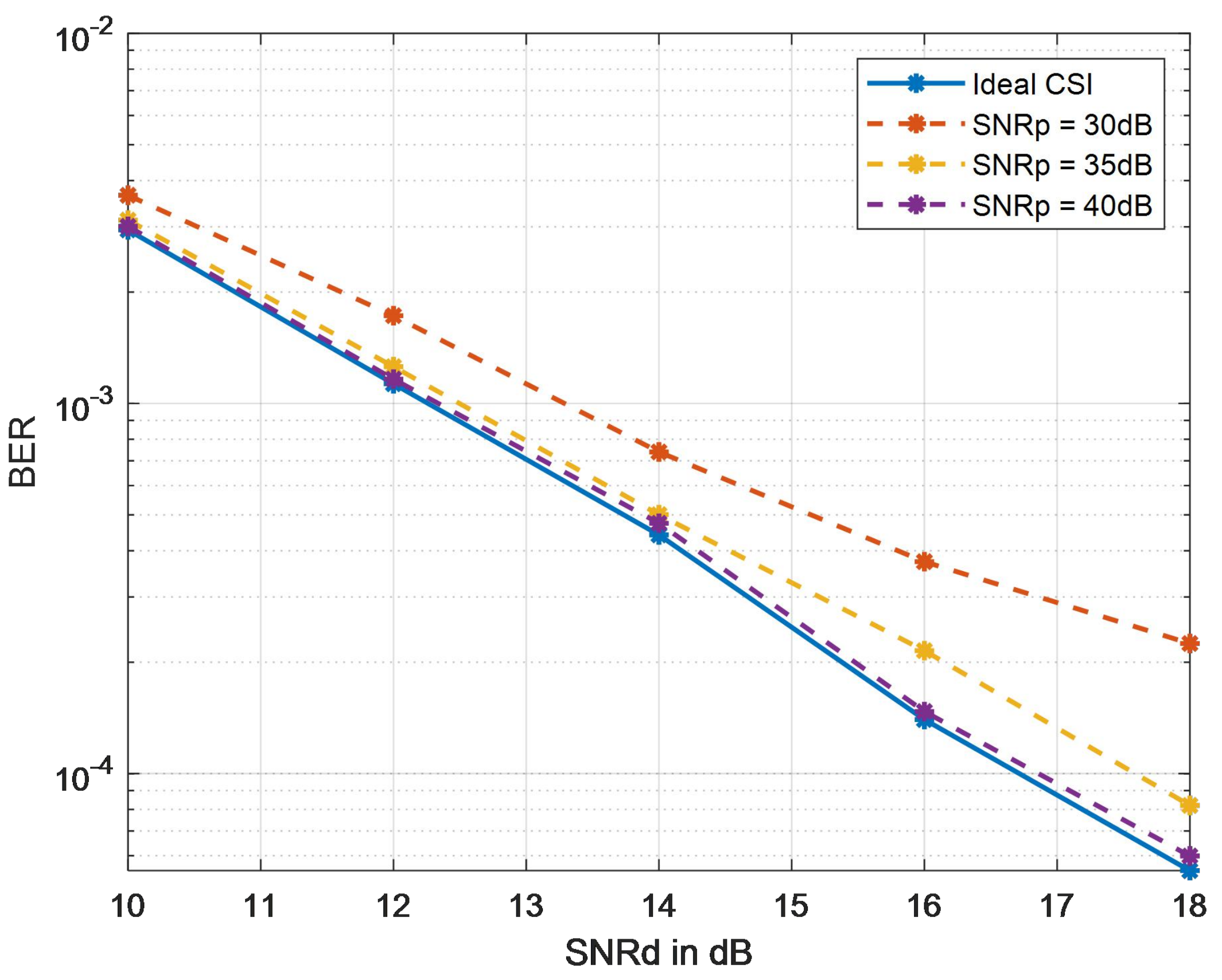}
	\caption{BER versus SNRd under different SNRp with optimal threshold for each case, $N$ = 256.}
	\label{4-3}
\end{figure}

Fig.\ref{4-3} shows the BER versus SNRd under different SNRp with optimal estimation threshold for each case. The SPA scheme is applied, and we can observe that the BER reduces as SNRp increases, which means a more accurate channel estimation is provided. We can also notice that when SNRp = 40dB, the BER performance of AFDM with estimated CSI is very close to AFDM with ideal CSI, which reveals the effectiveness of SPA scheme.

\begin{figure}[htbp]
	\centering
	\includegraphics[width=0.49\textwidth,height=0.39\textwidth]{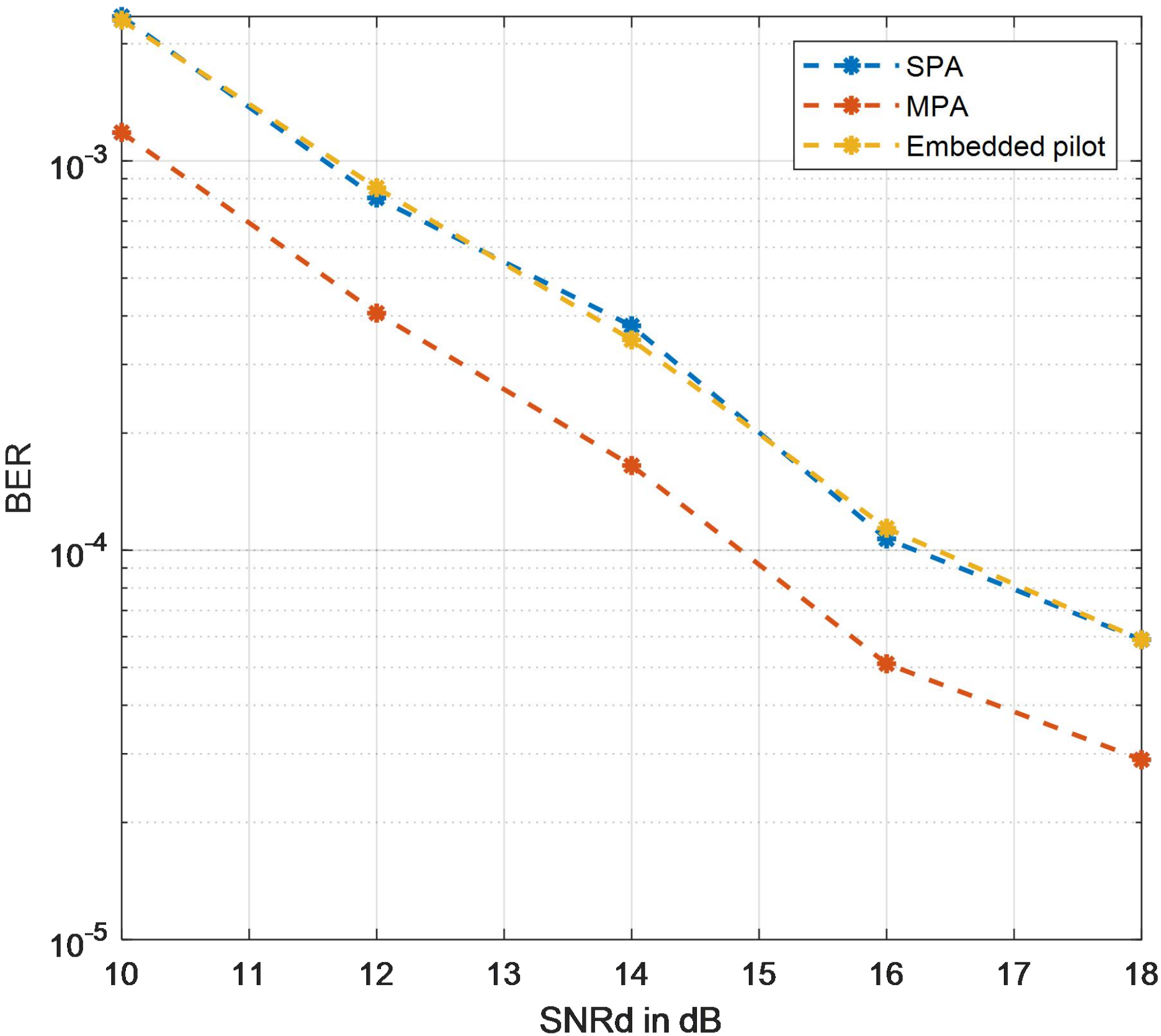}
	\caption{BER versus SNRd for different estimation schemes, SNRp = 20 dB, $N$=512.}
	\label{4-4}
\end{figure}

Fig.\ref{4-4} compares the embedded pilot, SPA and MPA scheme with SNRp = 20 dB. The estimation threshold $\zeta$ is set to 4 for each case and the number of pilots used in MPA is two. We can observe that the embedded pilot scheme has the same performance as SPA, proving the effectiveness of guard symbols arrangement in Section III. Moreover, with only one more pilot applied, the accuracy of channel estimation enhances distinctly since pilot diversity gain is explored.

\begin{figure}[htbp]
	\centering
	\includegraphics[width=0.49\textwidth,height=0.39\textwidth]{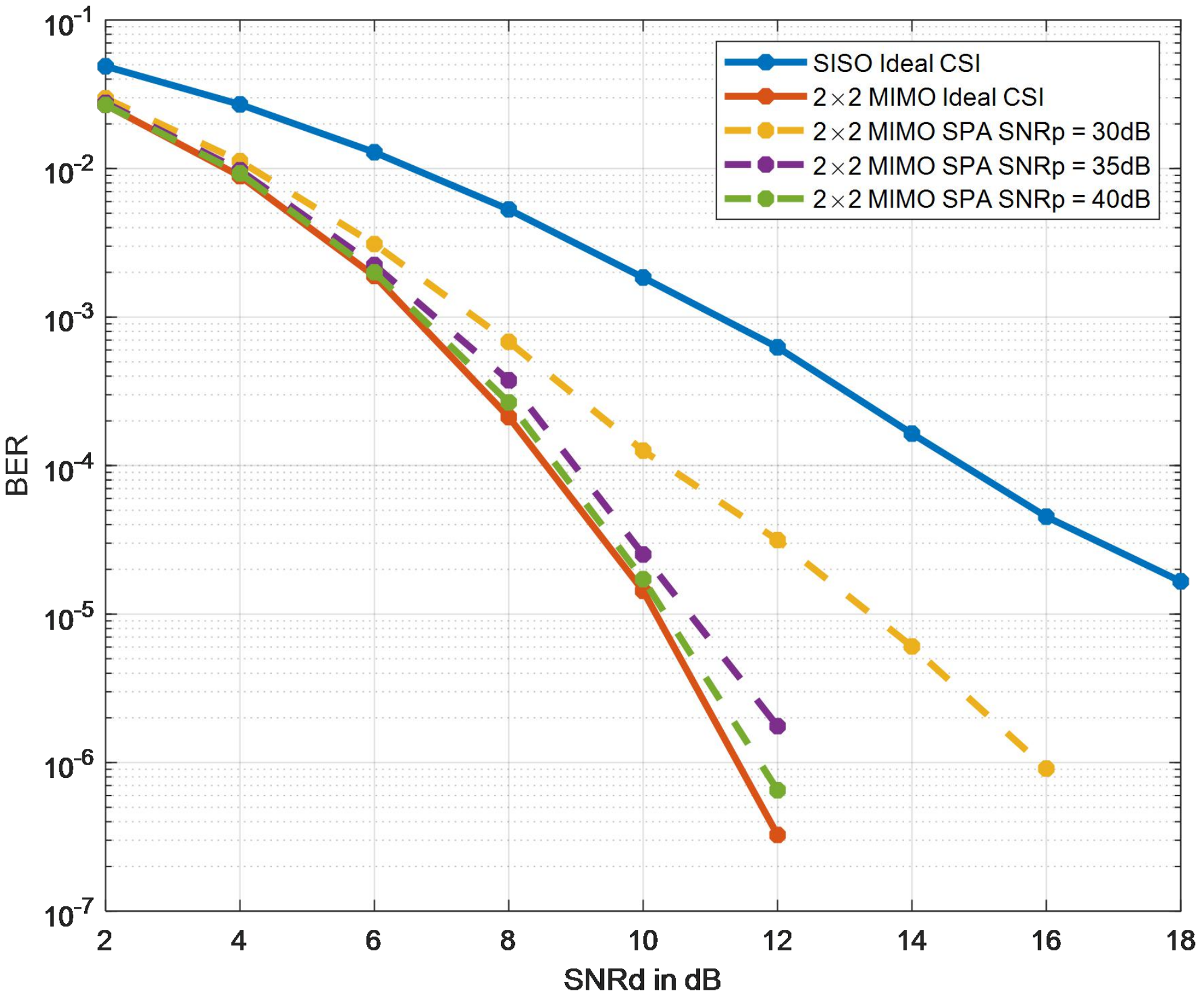}
	\caption{BER performance of $2\times2$ MIMO-AFDM system, $N$=512.}
	\label{4-5}
\end{figure}

Finally, Fig.\ref{4-5} shows the BER performance of $2\times2$ MIMO-AFDM system with ideal CSI and estimated CSI applying SPA scheme. The optimal estimation threshold is used for each non-ideal case and SISO-AFDM with ideal CSI is given as a reference. We can observe that in the case of SISO-AFDM with ideal CSI, SNRd = 18 dB is needed to approximately achieve the $10^{-5}$ BER, while only SNRd = 10 dB is needed in $2\times2$ MIMO-AFDM with ideal CSI. This enhancement results from space diversity gain provided by MIMO technology. We can also observe that when SNRp = 40 dB, the BER performance of MIMO-AFDM with estimated CSI is very close to MIMO-AFDM with ideal CSI, which verifies the effectiveness of SPA scheme in MIMO-AFDM system.

\section{Conclusion}
In this paper, we have proposed two pilot aided channel estimation schemes for AFDM. In particular, we arrange pilot, guard, and data symbols in the DAFT domain to suitably avoid interference between pilot and data symbols.  The influence of estimation threshold and pilot power on BER are discussed with simulations.  The results show that, by applying the proposed channel estimation scheme, AFDM with estimated CSI can achieve similar performance compared to AFDM with ideal CSI. 
Extensions of the SPA scheme to MIMO-AFDM and multi-user uplink/downlink have been presented, showing that AFDM maintains its advantage on OTFS in terms of overhead for pilot aided channel estimation in MIMO system.

\end{document}